\pdfoutput=1
\documentclass[12pt,journal,comsoc,twocolumn]{IEEEtran}
\usepackage[T1]{fontenc} 
\usepackage{amsmath}

\usepackage{bm} % optional
\usepackage{graphicx}
\ifCLASSOPTIONcompsoc
\usepackage[caption=false,font=normalsize,labelfont=sf,textfont=sf]{subfig}
\else
\usepackage[caption=false,font=footnotesize]{subfig}
\fi
\usepackage{url}
\usepackage{tikz}

\graphicspath{
	{images/}
}

\IEEEoverridecommandlockouts

\newcommand\copyrighttext{%
	\footnotesize \textcopyright 2019 IEEE. Personal use of this material is permitted.
	Permission from IEEE must be obtained for all other uses, in any current or future
	media, including reprinting/republishing this material for advertising or promotional
	purposes, creating new collective works, for resale or redistribution to servers or
	lists, or reuse of any copyrighted component of this work in other works.
	DOI: \href{https://doi.org/10.1109/MCOM.2019.1800949}{10.1109/MCOM.2019.1800949}}
\newcommand\copyrightnotice{%
	\begin{tikzpicture}[remember picture,overlay]
		\node[anchor=north,xshift=0pt,yshift=-10pt] at (current page.north) {\fbox{\parbox{\dimexpr\textwidth-\fboxsep-\fboxrule\relax}{\copyrighttext}}};
	\end{tikzpicture}%
}

\title{Delivering Gigabit Capacities to Passenger Trains\\Tales from an Operator on the Road to 5G}

\begin{document}
\newdimen\origiwspc%
\origiwspc=\fontdimen2\font%
\fontdimen2\font=0.86\origiwspc%

\bstctlcite{ShortCTL:BSTcontrol}
\author{
  \IEEEauthorblockN{Nima Jamaly, Stefan Mauron, Ruben Merz, Adrian Schumacher, Daniel Wenger\textsuperscript{\textdagger}}\\
  \IEEEauthorblockA{Swisscom  Ltd., Switzerland\\
    \small{\{nima.jamaly, stefan.mauron, ruben.merz, adrian.schumacher, daniel.wenger1\}@swisscom.com}}
  \thanks{\textsuperscript{\textdagger}Authors in alphabetical order.}
}
\maketitle

\copyrightnotice

% We use two terms: 1. (train) consist for composition; 2. railway car for a single wagon/carriage
% See http://www.electropedia.org/iev/iev.nsf/index?openform&part=811

\begin{abstract}
  Delivering reliable and high-capacity Internet connectivity to
  high-speed train users is a challenge. Modern railway cars act as
  Faraday cages and a typical train consist comprises several hundreds
  of users moving at high velocity. Furthermore, with the global
  availability of fourth generation (4G) Long Term Evolution (LTE),
  user expectations have dramatically increased: it is expected to be
  online anytime and anywhere. Demand for mobile high-capacity is
  being driven by video and music streaming services, for lower
  latency and higher availability by gaming, and for more reliability
  and even uplink capacity by mission critical applications. Finally,
  the life-cycle of the railway industry is much longer than for
  telecommunications, which makes supporting 5G challenging. In this
  paper, we survey the challenges associated with delivering
  high-capacity connectivity to train users, describe potential
  options, and highlight how a leading western European operator is
  tackling these challenges and preparing for 5G and beyond.
\end{abstract}

\section{Introduction}
\label{sec:introduction}

Delivering reliable mobile connectivity to train users has always been
challenging. First, railway cars with their metallic bodies and
low emissivity (Low-E)
coated glass windows for thermal isolation act as Faraday cages and cause high signal
attenuation that need compensation. For the typical frequency range of 800~MHz to 2600~MHz, measurements have shown
an attenuation of $25$ to $30$\,dB~\cite{burnier_energy_2017}.
Second, a train can transport several hundreds to
more than one thousand users. In a typical western European country,
this represents hundreds of Long Term Evolution (LTE)
active user-equipments (UE), traveling at speeds up to $330$\,km/h.
Advanced Doppler compensation algorithms are required, and network
dimensioning and optimization is non-trivial because the high rate of
handover and cell-reselection caused by high train velocities need to
be managed carefully: for example, cell-reselections or
handovers can take place every $40$ seconds for up to $1200$ users
simultaneously~\cite{ruben2014ltemeas}.

Finally, the deployment of 
4G has fueled an increasing demand for high-capacity mobile
connectivity. Mobile users expect unlimited access
to video and music streaming services, instant access to social media
services, and reliable connectivity for gaming services. These
expectations remain when users are commuting or traveling.
For instance, in Switzerland, with the densest railway network in
Europe, about 720\,km of railway tracks are used daily by more than
one million commuters. This puts a significant pressure on network and
train operators to provide adequate connectivity~\cite{ai2014}.
Train and network operators around the world are trialing next
generation solutions: for example in France~\cite{kaltenberger_broadband_2015}, in China~\cite{ai2014}, but also in Germany, UK, or Japan.\looseness=-1

Before describing our contribution, two more aspects need to be
discussed. The first is associated with the major strategy used by
service providers and railway operators to provide connectivity to
train users by deploying on-board equipment, for
instance, signal repeaters, or Wi-Fi routers backhauled via the cellular network.
This strategy is threatened in two ways: (1) The use of multiple-input multiple output
(MIMO) systems popularized in 4G and improved in fifth generation systems (5G) with higher order
MIMO, the use of both time-division duplexing (TDD) and
frequency-division duplexing (FDD) in 5G, and the use of
millimeter-wave spectrum in 5G, dramatically increasing system
complexity and cost for on-board equipment;
(2) upgrading this equipment to keep pace with standardization
progress is impeded by the life-cycle duration
difference between the railway industry and the telecommunications industry: 
typically, train infrastructure has life-cycles of tens of years,
whereas wireless communications infrastructure is upgraded every few
years (mobile network operators started LTE services late 2012 and
already upgraded to LTE Advanced and LTE Advanced Pro). To upgrade railway
infrastructure with the newest wireless communications infrastructure
every three to five years is for train operators a costly if not
impossible mission.
Another aspect is tied to 5G,
with which service providers will not only serve broadband
applications, but also -- enabled by network slicing -- future railway mobile
communication services (FRMCS)~\cite{Moreno_2015_FRMCS}.\looseness=-1

Hence, in this paper, our contribution is threefold. First, a survey
of the challenges associated with supporting high-capacity train
connectivity is presented in Section~\ref{sec:challenges}.
Then, a state of the art overview of how these
challenges are addressed by a majority of operators
is given in Section~\ref{sec:options}. However, this
current approach is threatened by the arrival of 5G. Finally, our third
contribution in Section~\ref{sec:solution} is a first look at how a leading western European operator
is
revisiting these approaches to tackle the foregoing challenges, and preparing for 5G.
Section~\ref{sec:conclusion} concludes the paper.

\section{What are the Challenges of Modern Railway Connectivity?}
\label{sec:challenges}

In this Section, we describe the most relevant challenges faced by
service providers and train operators to provide high-capacity
connectivity to train users.

\subsection{Railway Car Penetration Loss}

Modern railway car hulls are typically built of metal, creating a
Faraday cage. While the window apertures potentially allow signals to get
into the railway car, this effect is negated by the use of so-called
Low-E coated glass windows~\cite{burnier_energy_2017}.

\subsection{High-Velocity}

Intercity high-speed trains in Europe travel between $160$ and $330$~km/h.
In Asia similar velocities are used for commercial operation.
High velocity results in strong Doppler effects, creating various
performance degradations \cite{kaltenberger_broadband_2015,tse_fundamentals_2005}: Firstly, it creates
distortions on the signal at the receiver that need to be compensated.
Secondly, the Doppler effect is directly related to the
coherence time of the propagation channel. It is a measure of how
stable the channel remains over time. Fast evolving channels are more
difficult to estimate (e.g., averaging over time is limited) and result in
lower data rates. Last, a small coherence time highly decreases the
potential of feedback-based pre-coding schemes for MIMO systems.
High velocity also increases the rate of handovers and
cell-reselections~\cite{ruben2014ltemeas}.
If the network is not planned accordingly for large train consists (see below), this can
result in performance anomalies because of control-plane overload, temporary stalling at the
transport layer, and in the worst-case, increased
control-channel loss.

\subsection{Large Number of Co-Located Users}

In a cellular network, users share the available sector
bandwidth and control-plane capacity.
Trains cluster a large number of active terminals traveling simultaneously through
the network.
For instance, a typical $400$~meters long inter-city train consist can carry about $1200$ passengers during rush hours.
With trains crossing or at railway stations, even a larger number of active users per sector may have to be considered.
Such large numbers of users require operators to over-dimension their
infrastructure in order to not decrease service quality and
throughput.

\subsection{Supporting MIMO End-to-End}

MIMO is the key feature to support high-capacity connectivity via spatial multiplexing schemes, and
improve the overall robustness of the connectivity via diversity~\cite{tse_fundamentals_2005}.
Now, in case of a train scenario, with e.g., a signal repeater and radiating cables (or antennas) deployed onboard the railway car (see
Fig.~\ref{fig:leakyfeeder}), it is
critical that the entire RF chain supports MIMO. If not, it will mostly prevent the use of spatial multiplexing or diversity. While 2$\times$2 MIMO support has been validated \cite{inwagon2015mimo}, 4$\times$4
MIMO or massive MIMO schemes will be challenging. The latter is particularly valid in case
channel state information at the transmitter is required, or if more than two receive antennas are needed.

\subsection{Life-Cycle Industry Differences}

A major challenge for active solutions in railway cars is the difference between technology cycles of telecommunication infrastructure ($3$ to $5$ years) and the life-cycle of railway infrastructure ($15$ to $20$ years). It is not feasible and sometimes not possible to replace connectivity-related equipment in railway cars every couple of years. Indeed, railway cars traditionally require an in-depth revision every $15$ to $20$ years. Only during such events can extensive changes e.g., in cabling or communication equipment, be undertaken.

\subsection{Limited Space and Flexibility}
\label{sec:challenge_space}

Whichever active communications equipment is installed inside a railway
car, it will require one or several external antennas. Ideally, they are installed on the railway car's roof or side of the roof. However, depending on the railway car's size (considering double-deck railway cars), the clearance gauge does not allow for large antennas.

Furthermore, limited space, energy and air-conditioning along with an
extended operation temperature range poses challenges for active
communications equipment within railway cars. Cabling between
different components such as e.g., antennas and communications
equipment becomes very difficult (thermal insulation, railway car
structure, interior design), and is very limited between railway
cars.

\section{How are Challenges Addressed?}
\label{sec:options}

In this Section, we describe how challenges listed in
Section~\ref{sec:challenges} are typically addressed by service providers and train
operators.

\begin{figure}
	\centering
	\includegraphics[width=\linewidth]{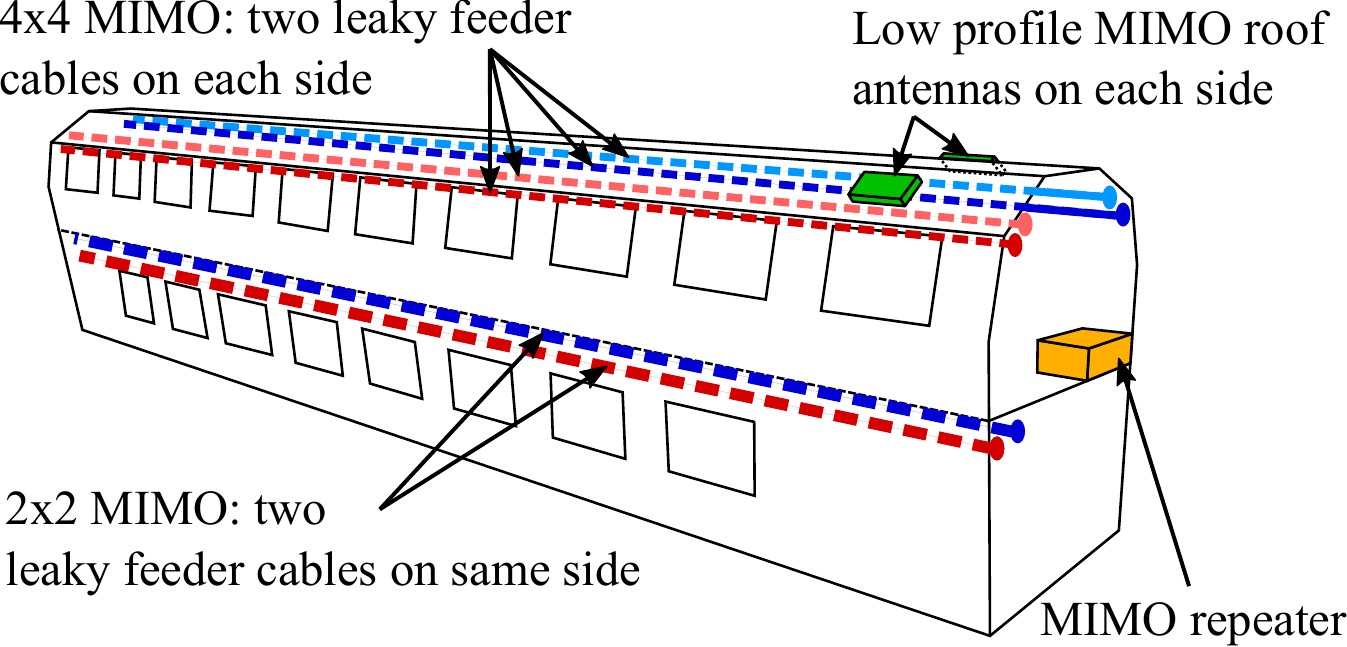}
	\vspace{-1cm}
	\caption{Railway car equipped with roof antennas, a repeater, and leaky
		feeder cables. Not shown are interconnecting jumper cables.\label{fig:repeaterwagon}}
\end{figure}

The most important challenge to address is to compensate the railway
car penetration loss, followed by -- in the context of 4G and 5G network
deployment -- the need to support end-to-end MIMO.
The most popular solution pursued to combat the penetration loss is by
outfitting the railway car with active equipment such as repeaters or
relays \cite{fokum2010survey}.
This implies the installation of outside roof antennas for the
backhaul (or donor link) and antennas or radiating cables (also known as leaky feeder cables, see~\cite{inwagon2015mimo}) inside the
railway car for
signal distribution (see Fig.~\ref{fig:repeaterwagon}).
We describe these options in more details in the following Section.

\subsection{Repeater, Relay Node and Moving Cell}
A repeater amplifies the input port signal and sends it out on the output port. 
While operating at the physical layer, modern repeaters are implemented in the digital domain.
Their main advantage is the independence of the radio access technology (RAT) and therefore transparency to wireless communications equipment.
However, in order to prevent feedback loops
and the resulting signal degradation or amplifier saturation, a repeater requires a high RF isolation between the inside radiating element(s) to
the outside antenna(s) (the amplified signal is on
the same frequency as the donor). The
current observed RF isolation of repeater equipped railway cars is between
$60$ to $80$\,dB.
Another well-known disadvantage is that repeater deployments can be responsible for large noise-floor elevation, which can severely limit uplink capacity \cite[Section~2.1]{ofcom_repeaters_2015}.
Furthermore, due to signal filtering in the repeater, a signal delay
(filter group delay) is introduced into the system.
Some repeaters allow to keep this delay below the length of the cyclic prefix of LTE to prevent interference between the inside and outside signals.
If the RF isolation is not sufficiently high, this signal delay
together with the ratio of repeater gain to
railway car RF isolation causes unwanted self-interference.

A relay-node (RN) is another option, similar to repeaters, but operating on the upper layers.
An RN can solve some of the problems observed with repeaters.
Due to signal decoding and re-encoding, the re-transmitted signal is
clean of noise and possible interference. Also, numerous users can
be aggregated to significantly save on control-plane signaling, especially
during handovers.
However, RNs also require a high 
system isolation in case the same bands are
used. Further disadvantages are the dependency on RAT releases and feature support, resulting in RN
firmware or hardware upgrades, and the use of more
space and electrical power.
From a 3GPP point of view, RNs are only standardized for LTE and without mobility support.
To the best of our knowledge, today no LTE RN is in operation in the railway context.\looseness=-1

A further evolution of the RN concept is the so-called moving-cell,
where UEs on the mobile link do not experience handovers or
cell-reselection. Only the donor link handles these events. The
moving-cell can help solving the challenge related to a large number of
users. However, this implies that a large part of the protocol stack
of the underlying air-interface(s) be implemented in the active
equipment. Because typical deployments today need to support multiple
RAT generations, from 2G to 5G, we are not aware of any moving cell in
operation today.

Common for repeaters and RNs is the problem with concurrent operation
of TDD and FDD links on the same RF transmission systems, which
dramatically increases the complexity of isolation and filtering.

\subsection{On-Board Wi-Fi}

Another way of providing wireless connectivity inside railway cars is by
means of an on-board Wi-Fi access point. The backhaul link of course uses cellular connectivity, but the mobile link is using Wi-Fi.
Contrary to repeater or RN
equipment, such a system is much more independent of the railway car hull
isolation, because different frequency bands are used. Although care needs to be taken between the ISM band at $2.4$\,GHz and LTE bands 7 or 38 around $2.6$\,GHz.
Regarding the deployment of Wi-Fi inside railway cars, it should be noted
that a performance-oriented system must support the UNII $5$\,GHz band and
recent standards (IEEE 802.11ac or 802.11ax).

Deployments of mobile routers with Wi-Fi in the railway cars may prove more
advantageous from a cost, power consumption, and evolution point of
view. However, they do not solve the need for a high-quality backhaul,
antennas on the roof, or radiating cables
or antennas inside the railway car.
While it is possible to use Wi-Fi calling, it may not work for all
passengers, and an end-to-end quality of service control is also not
straightforward to achieve. 
Additionally, because many users make heavy use of Wi-Fi tethering,
the $2.4$ and $5$\,GHz bands are already fairly crowded.

\subsection{Donor Link: MIMO Roof Antennas}

\begin{figure}
	\centering
	\subfloat[\label{fig:HSlowprofile}\vspace{-1em}]{%
		\includegraphics[width=0.49\linewidth]{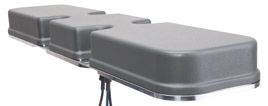}}
	\hfill
	\subfloat[\label{fig:leakyfeeder}\vspace{-1em}]{%
		\includegraphics[width=0.49\linewidth]{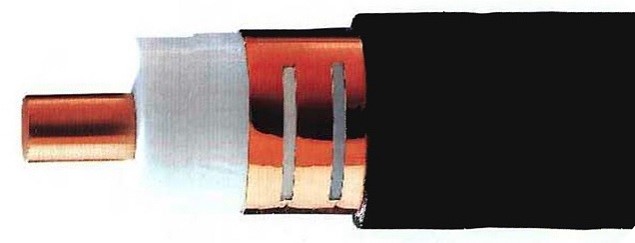}}
	\caption{(a) Low-profile 2$\times$2 MIMO roof antenna for $790$-$2700$\,MHz and
		$4.9$-$6.4$\,GHz. (b) Example of a radiating cable.}
\end{figure}

All active communications equipments discussed so-far
require donor link antennas, ideally installed on the railway car roof.
As discussed in the previous Section~\ref{sec:challenge_space}, the clearance gauge
may not allow for large antennas on the roof.
For example, to solve this challenge for train operators, we developed a low-profile MIMO antenna together
with Huber+Suhner for double-deck railway cars where a maximum height of 
$40$\,mm is possible, see Fig.~\ref{fig:HSlowprofile}.

At least two antenna elements are required to support 2$\times$2 MIMO.
For higher order MIMO or diversity, more antenna elements need to be installed. Furthermore,
there are very often constraints regarding the installation locations on the roof,
available installation space, and orientation.
Also, care must be taken to ensure that the rooftop antenna elements' position allows for a high-rank propagation channel.

\subsection{Propagation Inside Railway Cars: MIMO Radiating Cables}

Inside railway cars, radiating elements also need to be deployed.
Radiating cables (see Fig.~\ref{fig:leakyfeeder})
are typically used: they are easy to install, along the length of
the railway car interior, either under the ceiling, along the side, or under
the floor; radiating cables also create a more uniform
signal along the railway car compared to antenna deployments; finally, radiating cables can be easily hidden from passenger view. Indeed, antenna deployments have sometimes created controversies with passengers uncomfortable with radio-wave deployments.

State of the art radiating cables support frequencies in
the range of a few hundred MHz up to $2600$\,MHz. However, more wideband
support on a single cable appears quite challenging. Support of both,
cellular frequencies and the ISM band at 5\,GHz, would require the
deployment of two-cables. Deployment of MIMO spatial multiplexing
requires additional cables as well, which has been validated
with extensive measurements. In~\cite{inwagon2015mimo} we demonstrated
that 2$\times$2 MIMO as well as 4$\times$4 MIMO can be supported.

\subsection{What are Most Operators Using Today?}

The majority of operational deployments today are building on
repeaters or on-board Wi-Fi, or a combination of both. Antennas or
radiating cables are used within railway cars. And while solutions are
available for supporting 2$\times$2 MIMO end-to-end, they are not widely deployed
because of life-cycle challenges.

Indeed, the life-cycle challenge is not solved at all by current
solutions, and its importance will further grow at the advent of 5G
and the evolution of Global System for Mobile Communications – Railway (GSM-R)~\cite{uic_gsm-r}
to FRMCS \cite{Moreno_2015_FRMCS}.
Hence, before moving on to the next Section where we briefly describe 
these challenges, we will in short outline some of the relevant
specifics of 5G and FRMCS.

\subsection{The Rise of 5G}
New Radio (NR) is the successor RAT of LTE currently
being specified by 3GPP in Release 15 and 16. The first commercial network
deployments have recently been switched on and early devices are
expected to appear during the first half of 2019. With ultra lean transmission,
latency optimized frame structure, massive MIMO and interworking
between low and very high frequency bands (from 600\,MHz up to
86\,GHz \cite{lee_spectrum_2018})
it will dramatically increase data-rates available to train passengers and applications.
The latency improvements, the
support of higher-reliability and mission-critical applications, and
the availability of network slicing will also  be beneficial for
train applications, especially in the context of FRMCS applications.

Deploying NR in millimeter-wave (mmWave) frequencies (e.g., $f > 24$\,GHz) mandates the use of antenna
arrays with e.g., 64-elements or more, adaptive
beamforming, and TDD. Concurrently, a network densification is required due to
increased free-space path loss and decreased signal penetration through
obstacles.

In Europe, a focus on NR deployments at $3.5$-$3.8$\,GHz, using carrier bandwidths of $100$\,MHz or more, TDD, and antenna arrays, followed by mmWave deployments is expected.

\subsection{Future Railway Mobile Communication System}
\label{sec:frmcs}
Focusing on Europe, currently many operators rely on their own GSM-R networks.
They are primarily used to transport voice communication
(point-to-point, group, broadcast and emergency calls) and most importantly, data
traffic of the European Train Control System (ETCS) \cite{winter_etcs_1992}. Today, ETCS Level 2
is in use, but the change to level 3 is being evaluated. ETCS level 3
requires accurate and reliable positioning and relies on radio-based train spacing.
This will support increasing the capacity of heavily loaded railway networks. Hence, many
train operators plan a thorough digitization of end-to-end railway
operations, with an outlook on autonomous train operation.
Future railway operation
demands for higher data rates and therefore new spectrum and a network
densification leveraging synergies of the telecommunications industry.
Thus, the railway industry, aligning with the telecommunications
industry, is developing FRMCS in order to support this digitalization effort, and the eventual decommissioning of GSM-R.
The first FRMCS trial implementations are expected to start around 2020.
Therefore, in addition to provide mobile connectivity to train users,
service providers will also likely have to consider the support of
FRMCS services.

\section{Bringing High-capacity and Supporting 5G by Decoupling Life-cycles}
\label{sec:solution}

While on-board repeaters or Wi-Fi routers have become popular to
deliver capacity for railway users, they have also become a
bottleneck.
Indeed, on-board equipment is responsible for the life-cycle coupling
challenge. With the advent of 5G, bringing spectrum at $3.5$ GHz, a
TDD air-interface and massive MIMO, a major equipment upgrade is
necessary. Furthermore, 5G with its dual use of TDD and FDD,
dramatically increases the requirements on on-board equipment.

Therefore, in order to reduce the life-cycle
coupling challenge as much as possible, and to ensure that railway users can readily
benefit from cellular infrastructure innovation, we have decided to
move away from using on-board equipment.
Instead, we have decided to promote and use so-called RF transparent
Low-E windows~\cite{burnier_energy_2017} (hereafter RF transparent
windows).

\subsection{RF Transparent Windows make On-board Equipment Obsolete}
In modern railway cars,  the window panes have a thin metallic coated layer for enhanced thermal (infrared) insulation. However, such a metallic coating increases the RF 
transmission loss through them significantly (e.g., $25-30$\,dB~\cite{burnier_energy_2017}). Recently, by adjusting the coating pattern on the window panes, engineers have successfully reduced the transmission loss for sub $3$\,GHz bands to only a couple of dBs. 
Simulations and measurements of railway cars equipped with RF transparent windows in real conditions demonstrated a reduction of around $12-14$\,dB at $1.8$\,GHz band~\cite{burnier_energy_2017, jamaly_analysis_2018}. 
Using frequency selective surface concepts in electromagnetics, scientists are expected to come up with an improved coating pattern which could likewise serve to reduce the transmission loss for frequency ranges up to $6$\,GHz.
As such, classic 2$\times$2 or higher order MIMO spatial multiplexing is
supported.

RF transparent windows are the crucial parts for supporting
high-capacity for railway applications. In addition, earlier studies,
for example~\cite{ruben2014ltemeas}, have clearly shown that even at
the high-velocity characteristics of railway scenarios, ensuring a
high signal to interference and noise ratio (SINR) is the key factor
to benefit from high spectral efficiency offered by spatial
multiplexing schemes. As such, we are also promoting the use of
dedicated cellular infrastructure for railway coverage. This design
choice is also guided by the upcoming requirements of FRMCS (see
Section~\ref{sec:frmcs}).

\subsection{Dedicated Infrastructure for high SINR: the RF Corridor}

In order to increase SINR, a dedicated infrastructure is required, deployed as
close as possible to the railway track, a so-called RF corridor. Indeed, tight
emmission regulations 
prevent increasing existing
sector power.
A dedicated infrastructure is a form of
sectorization and densification, which helps supporting large number of users. A dedicated infrastructure also enables specific
optimization for railway scenarios, but also improved localization
support (e.g., for FRMCS). Combined sectors can also be deployed to reduce
cell-reselection and handover.

Distributed antenna systems (DAS) or cloud-RAN approaches can be used
to build such infrastructure. DAS are favored in case of
multi-operator support requirements. However, they often lack
proper (digital) interworking with base-station equipment to support the use of
active antenna mechanisms or sophisticated MIMO algorithms.

An RF corridor may need integration in the existing
macro-network. Various options are available, from using dedicated
bands, to benefiting from improved spatial reuse by beamforming
(possible with NR) or heterogeneous network algorithms.

\subsection{Reflecting Panels}
\begin{figure}
	\centering
	\includegraphics[width=0.7\linewidth]{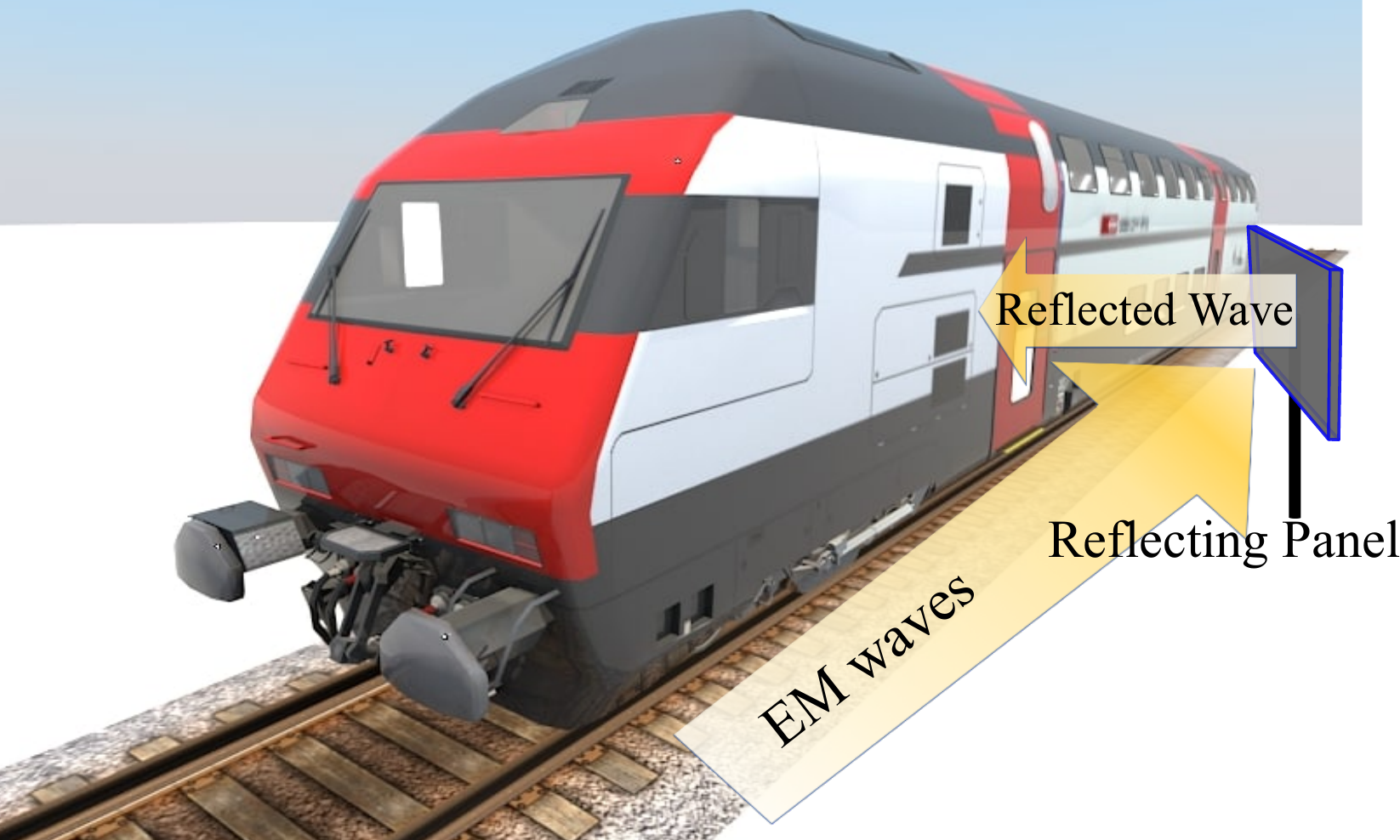}
	\vspace{-1em}
	\caption{Outlook of a reflecting panel in front of a railway car. \label{fig:panel}}
\end{figure}

Our early simulations of RF corridor with RF transparent window-equipped trains have highlighted one potential issue:
in case antennas are installed near the railroad, wide angles of incident wave to the window panes become unavoidable, especially in scenarios with poor scattering and reflections provided by the surrounding environment. 
This configuration can increase the penetration loss considerably
(because of the significantly reduced cross-section area).
To combat this
performance anomaly, reflecting panels can be deployed along the railway tracks~\cite{jamaly_analysis_2019}: these conductive panels
indirectly increase the cross section area of the window panes from the
antenna standpoint. Fig.~\ref{fig:panel} illustrates such a scenario, where a simple flat conductive panel with $45^\circ$ angle towards the railway car is used to reflect RF energy towards the window panes.
The exact gain of a reflecting panel on path loss
strongly depends on its physical size and design. 

\begin{figure}
	\centering
	\includegraphics[width=\linewidth]{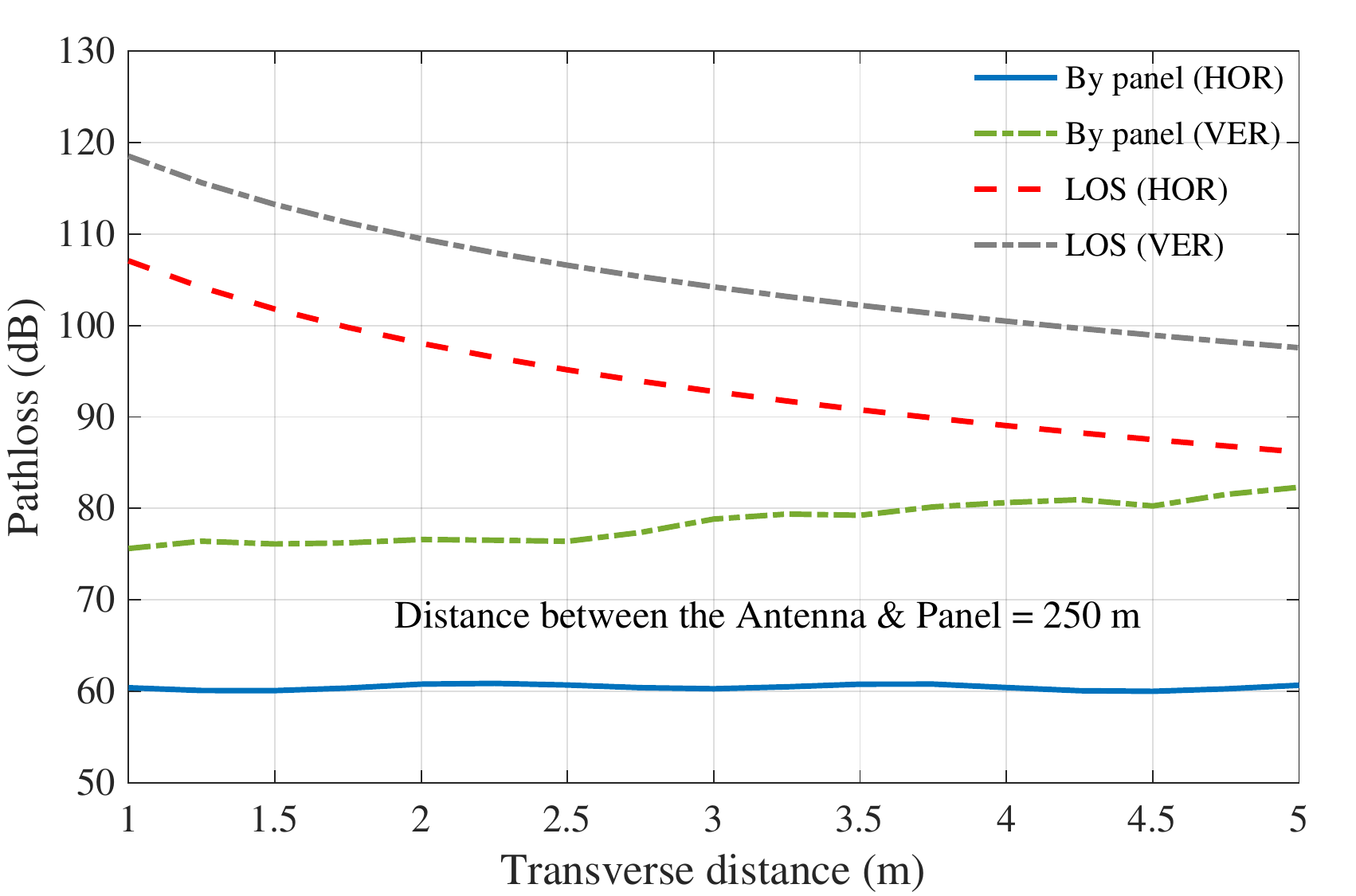}
	\vspace{-2em}
	\caption{Pathloss for the two polarizations with and without reflecting panel in an RF corridor scenario at $2.6$\,GHz. Gains of antennas are $\mbox{G}_{\mbox{tx}} = 25$\,dBi \& $\mbox{G}_{\mbox{rx}} = 0$\,dBi.}
	\label{fig:pathloss} 
\end{figure}

In Fig.~\ref{fig:pathloss}, we show the path loss between the ports
of a base station antenna (with 20\,dBi gain), and an ideal dual-port
omni-directional reference antenna (with 0\,dBi gain) on-board a typical
railway car. The illustrated path losses at $2.6$\,GHz are shown for
vertical (VER) and horizontal (HOR) polarizations (with respect to
the ground) and 
with and without a reflecting panel at a
distance of $250$\,m from the antenna. The window panes
used here are similar to those studied in~\cite{jamaly_analysis_2019} 
and scenarios without panel are referred to as line of sight (LOS).

Clearly, the impact of panels on the path loss is significant and rather independent of the traverse distance, as long as the traverse distance is a small fraction of the distance between the panel and the base station antenna.

\subsection{Outlook on Performance and Long-Term Spectrum Evolution}

We have already validated the benefit of RF windows in a regular
macro-network~\cite{burnier_energy_2017}. Throughout $2019$, we will
evaluate the improvements brought by deploying an RF corridor
supported by reflecting panels.
Early simulation results demonstrate that average spectral
efficiencies can be doubled compared to regular macro-network
coverage.

Looking forward, while many initial 5G deployments will begin with $3.5$~GHz, systems in mmWave frequencies will
eventually be deployed. To allow the use of such frequency ranges for
train passengers, we expect to build on the existing RF transparent
windows. Indeed, while RF transparent windows do not completely remove
the life-cycle challenge, they reduce it significantly. Exchanging windows is much easier than a complete on-board RF installation.
And recent demonstrations such as mmWave on-glass antennas are
promising~\cite{agc2018}.

\section{Conclusion}
\label{sec:conclusion}

In this paper, we have reported on how a leading western European operator
intends to evolve railway cars with RF transparent windows, and combine this
approach with RF corridors and reflective panels to deliver very high
capacity connectivity to train customers. This combination will allow
for the support of 5G deployments up to $3.5$\,GHz without any
complex and costly on-board equipment.
RF transparent windows have already been validated in the field. Their
validation with an RF corridor and reflecting panels, in the context
of early 5G deployments is ongoing.
The use of mmWave for trains will also be addressed.

\bibliographystyle{IEEEtran}
\bibliography{IEEEabrv,articleBib}

% Generated by IEEEtran.bst, version: 1.12 (2007/01/11)
\begin{thebibliography}{10}
\providecommand{\url}[1]{#1}
\csname url@samestyle\endcsname
\providecommand{\newblock}{\relax}
\providecommand{\bibinfo}[2]{#2}
\providecommand{\BIBentrySTDinterwordspacing}{\spaceskip=0pt\relax}
\providecommand{\BIBentryALTinterwordstretchfactor}{4}
\providecommand{\BIBentryALTinterwordspacing}{\spaceskip=\fontdimen2\font plus
\BIBentryALTinterwordstretchfactor\fontdimen3\font minus
  \fontdimen4\font\relax}
\providecommand{\BIBforeignlanguage}[2]{{%
\expandafter\ifx\csname l@#1\endcsname\relax
\typeout{** WARNING: IEEEtran.bst: No hyphenation pattern has been}%
\typeout{** loaded for the language `#1'. Using the pattern for}%
\typeout{** the default language instead.}%
\else
\language=\csname l@#1\endcsname
\fi
#2}}
\providecommand{\BIBdecl}{\relax}
\BIBdecl

\bibitem{burnier_energy_2017}
L.~Burnier \emph{et~al.}, ``\BIBforeignlanguage{en}{Energy saving glazing with
  a wide band-pass {FSS} allowing mobile communication: up-scaling and
  characterisation},'' \emph{\BIBforeignlanguage{en}{IET Microwaves, Antennas
  \& Propagation}}, vol.~11, no.~10, pp. 1449--1455, Aug. 2017.

\bibitem{ruben2014ltemeas}
R.~Merz \emph{et~al.}, ``Performance of {LTE} in a {High}-velocity
  {Environment}: {A} {Measurement} {Study},'' in \emph{Proceedings of the 4th
  {Workshop} on {All} {Things} {Cellular}}.\hskip 1em plus 0.5em minus
  0.4em\relax New York, NY, USA: ACM, 2014, pp. 47--52.

\bibitem{ai2014}
B.~{Ai} \emph{et~al.}, ``Challenges toward wireless communications for
  high-speed railway,'' \emph{IEEE Transactions on Intelligent Transportation
  Systems}, vol.~15, no.~5, pp. 2143--2158, Oct 2014.

\bibitem{kaltenberger_broadband_2015}
F.~Kaltenberger \emph{et~al.}, ``Broadband wireless channel measurements for
  high speed trains,'' in \emph{2015 {IEEE} {ICC}}, Jun. 2015, pp. 2620--2625.

\bibitem{Moreno_2015_FRMCS}
J.~{Moreno} \emph{et~al.}, ``A survey on future railway radio communications
  services: challenges and opportunities,'' \emph{IEEE Communications
  Magazine}, vol.~53, no.~10, pp. 62--68, Oct 2015.

\bibitem{tse_fundamentals_2005}
D.~Tse and P.~Viswanath, \emph{\BIBforeignlanguage{en}{Fundamentals of
  {Wireless} {Communication}}}, May 2005.

\bibitem{inwagon2015mimo}
R.~Merz \emph{et~al.}, ``A {Measurement} {Study} of {MIMO} {Support} with
  {Radiating} {Cables} in {Passenger} {Rail} {Cars},'' in \emph{{IEEE} {VTC}
  {Spring}}, Glasgow, Scotland, May 2015, pp. 1--5.

\bibitem{fokum2010survey}
D.~T. Fokum and V.~S. Frost, ``A {Survey} on {Methods} for {Broadband}
  {Internet} {Access} on {Trains},'' \emph{IEEE Communications Surveys
  Tutorials}, vol.~12, no.~2, pp. 171--185, Second 2010.

\bibitem{ofcom_repeaters_2015}
``An {Assessment} of the {Effects} of {Repeaters} on {Mobile} {Networks},''
  Ofcom UK, Tech. Rep. GTC-15-0007-D, Nov. 2015.

\bibitem{uic_gsm-r}
\BIBentryALTinterwordspacing
``{GSM}-{R} - {UIC} - {International} union of railways.'' [Online]. Available:
  \url{https://uic.org/gsm-r} [Accessed: 2019-03-18]
\BIBentrySTDinterwordspacing

\bibitem{lee_spectrum_2018}
J.~Lee \emph{et~al.}, ``Spectrum for {5G}: {Global} {Status}, {Challenges}, and
  {Enabling} {Technologies},'' \emph{IEEE Communications Magazine}, vol.~56,
  no.~3, pp. 12--18, Mar. 2018.

\bibitem{winter_etcs_1992}
P.~Winter, ``The {Project} for a {Unified} {European} {Train} {Control} and
  {Protection} {System} ({ETCS}),'' \emph{Rail International}, no. 6-7, 1992.

\bibitem{jamaly_analysis_2018}
N.~Jamaly \emph{et~al.}, ``\BIBforeignlanguage{en}{Analysis and {Measurement}
  of {Penetration} {Loss} for {Train} {Wagons} with {Coated} vs {Uncoated}
  {Windows}},'' in \emph{\BIBforeignlanguage{en}{{EuCAP}}}, Apr. 2018, p. 610
  (5 pp.).

\bibitem{jamaly_analysis_2019}
N.~Jamaly, S.~Mauron, and A.~Kishk, ``\BIBforeignlanguage{en}{Application of
  {Reflecting} {Panels} in {Realisation} of {Antenna} {Corridor} for {Train}
  {Communications}},'' in \emph{\BIBforeignlanguage{en}{{EuCAP}}}, Apr. 2019,
  pp. 1--5.

\bibitem{agc2018}
\BIBentryALTinterwordspacing
AGC, ``Success with {5G} {Communications} {Using} ''{Vehicle} {Glass} {Mounted}
  {Antenna}'' for {5G} {Connected} {Car},'' July 2018. [Online]. Available:
  \url{http://www.agc.com/en/news/detail/1197413_2814.html} [Accessed:
  2018-12-14]
\BIBentrySTDinterwordspacing

\end{thebibliography}

\vspace*{-2\baselineskip}
\begin{IEEEbiographynophoto}{Nima Jamaly}
is a senior research engineer in Swisscom Innovations, Switzerland.  During the last few years, he has also been a guest research scholar in Chalmers University of Technology, Sweden. His main areas of research are antennas and propagation.
\end{IEEEbiographynophoto}
\vskip -2\baselineskip plus -1fil

\begin{IEEEbiographynophoto}{Stefan Mauron} 
holds an engineering degree in information-systems 
technology and in business administration, both from the 
University of Applied Sciences in Bern. Stefan is a principal product innovation lead for 
wireless access for the 5G program at Swisscom.
\end{IEEEbiographynophoto}
\vskip -2\baselineskip plus -1fil

\begin{IEEEbiographynophoto}{Ruben Merz}
  is the lead system architect for the 5G
  program at Swisscom. 
  Ruben earned his PhD and MSc from the
  School of Computer and Communication Systems at EPFL, and a master
  of advanced studies in management, technology \& economics at ETH
  Z\"urich
  
\end{IEEEbiographynophoto}
\vskip -2\baselineskip plus -1fil

\begin{IEEEbiographynophoto}{Adrian Schumacher} [M'07]
received his M.Sc. degree from the Royal Institute of Technology (KTH), Stockholm. He is a senior engineer at Swisscom Innovations.
His research interests include digital signal processing, wireless communications, and mmWave.
\end{IEEEbiographynophoto}
\vskip -2\baselineskip plus -1fil

\begin{IEEEbiographynophoto}{Daniel Wenger}
received his B.Sc. degree in Electrical Engineering from University of Applied Science in Bern (CH). He is a senior innovation engineer at Swisscom working in the 5G program.

\end{IEEEbiographynophoto}

\end{document}